%% file: 00main.tex
\documentclass[aps,prl,reprint,amsmath,amssymb,superscriptaddress, nofootinbib,floatfix]{revtex4-1}
\usepackage[utf8]{inputenc}
\usepackage[normalem]{ulem}
\usepackage[dvips]{graphicx}
\usepackage{dcolumn}
\usepackage{bm}
\usepackage[mathlines]{lineno}
\usepackage{lipsum}
\usepackage{hyperref}
\usepackage{xspace}
\usepackage[nolist]{acronym}
\usepackage{color}
\usepackage{subfigure}
\usepackage{enumitem}
\usepackage[usenames,dvipsnames,svgnames,table]{xcolor}	
\usepackage{verbatim}
\graphicspath{{images/}}

\input{comment_commands} 
\input{shortcuts}

\begin{document}

\title{Measurement of the \texorpdfstring{2\nbb}{2vbb} Decay Half-life of \texorpdfstring{$^{130}$Te}{130Te} with CUORE}

\input{author_list_2020_Oct_30}

\date{\today}

\begin{abstract}
We measured two-neutrino double beta decay of \isoTe using an exposure of \exposure\kgyr accumulated with the CUORE detector. Using a Bayesian analysis to fit simulated spectra to experimental data, it was possible to disentangle all the major background sources and precisely measure the two-neutrino contribution. The half-life is in agreement with past measurements with a strongly reduced uncertainty: \hlifeResult. This measurement is the most precise determination of the \isoTe 2\nbb decay half-life to date.
\end{abstract}
\maketitle

\input{acronyms}

\input{01intro}
\input{02detector}
\input{03data}
\input{04specfit}
\input{05bmsystematics}
\input{06results}
\input{07conclusion}
\input{08acknowledgments.tex}

\bibliography{09biblio}

\clearpage
\appendix

\input{10suppl}

\end{document}

%% file: comment_commands.tex
\newcommand{\maura}[1]{{\color{Purple}{#1}}}

\newcommand{\updateValue}[1]{{\color{Black}{#1}}}

%% file: shortcuts.tex
\newcommand{\kgyr}{kg$\cdot$yr\xspace}

\newcommand{\TeO}{TeO$_2$\xspace}
\newcommand{\isoTe}{$^{130}$Te\xspace}
\newcommand{\NDBD}{$0\nu\beta\beta$ decay\xspace}

\newcommand{\nbb}{$\nu\beta\beta$\xspace}
\newcommand{\vbb}{$\nu\beta\beta$\xspace}
\newcommand{\Qbb}{$Q_{\beta\beta}$\xspace}
\newcommand{\qino}{Cuoricino\xspace}
\newcommand{\q}{\mbox{CUORE-0}\xspace}
\newcommand{\GEANT}{G{\sc eant}}

\newcommand{\Tedecay}{\mbox{$^{130}\mathrm{Te} \to ^{130}\mathrm{Xe} + 2e^{-}$}\xspace}
\def\chisq{$\chi^{2}$\xspace}

\def\multi{$\mathcal{M}$\xspace}
\def\mone{$\mathcal{M}_1$\xspace}

\def\mtwo{$\mathcal{M}_2$\xspace}

\def\msumtwo{$\Sigma_2$\xspace}
\def\thalf{$T_{1/2}^{2\nu}$\xspace}

\def\datasets{\updateValue{7}\xspace}

\def\exposure{\updateValue{300.7}\xspace}
\def\isoexposure{\updateValue{102.7}\xspace}

\def\numSim{\updateValue{62}\xspace}
\def\numElements{\updateValue{9}\xspace}
\def\muonMulti{\updateValue{5}\xspace}
\def\dtCoinc{\updateValue{30~ms}\xspace}
\def\drCoinc{\updateValue{15~cm}\xspace}
\def\dECoinc{\updateValue{70~keV}\xspace}
\def\fitElo{\updateValue{350~keV}\xspace}
\def\fitEhi{\updateValue{2.8~MeV}\xspace}
\def\minBin{\updateValue{15~keV}\xspace}
\def\minCounts{\updateValue{30}\xspace}
\def\numBgListSyst{\updateValue{25}\xspace}
\def\chiglobal{\updateValue{681}\xspace}
\def\dofglobal{\updateValue{365}\xspace}
\def\redchiglobal{\updateValue{1.87}\xspace}

\def\hlife{\updateValue{7.71}}
\def\hlifeNoSr{\updateValue{7.67}}
\def\hlifestatUp{\updateValue{+0.08}}
\def\hlifestatLow{\updateValue{-0.06}}
\def\hlifestatNoSr{\updateValue{0.05}}

\def\hlifesystLow{\updateValue{-0.15}}
\def\hlifesystUp{\updateValue{+0.12}}
\def\hlifefracuncert{\updateValue{2.0}}
\newcommand{\hlifeResult}{\mbox{\thalf = $\hlife^{\hlifestatUp}_{\hlifestatLow}\mathrm{(stat.)}^{\hlifesystUp}_{\hlifesystLow}\mathrm{(syst.)}\times10^{20}$~yr}\xspace}

%% file: author_list_2020_Oct_30.tex
\author{D.~Q.~Adams}
\affiliation{Department of Physics and Astronomy, University of South Carolina, Columbia, SC 29208, USA}

\author{C.~Alduino}
\affiliation{Department of Physics and Astronomy, University of South Carolina, Columbia, SC 29208, USA}

\author{K.~Alfonso}
\affiliation{Department of Physics and Astronomy, University of California, Los Angeles, CA 90095, USA}

\author{F.~T.~Avignone~III}
\affiliation{Department of Physics and Astronomy, University of South Carolina, Columbia, SC 29208, USA}

\author{O.~Azzolini}
\affiliation{INFN -- Laboratori Nazionali di Legnaro, Legnaro (Padova) I-35020, Italy}

\author{G.~Bari}
\affiliation{INFN -- Sezione di Bologna, Bologna I-40127, Italy}

\author{F.~Bellini}
\affiliation{Dipartimento di Fisica, Sapienza Universit\`{a} di Roma, Roma I-00185, Italy}
\affiliation{INFN -- Sezione di Roma, Roma I-00185, Italy}

\author{G.~Benato}
\affiliation{INFN -- Laboratori Nazionali del Gran Sasso, Assergi (L'Aquila) I-67100, Italy}

\author{M.~Biassoni}
\affiliation{INFN -- Sezione di Milano Bicocca, Milano I-20126, Italy}

\author{A.~Branca}
\affiliation{Dipartimento di Fisica, Universit\`{a} di Milano-Bicocca, Milano I-20126, Italy}
\affiliation{INFN -- Sezione di Milano Bicocca, Milano I-20126, Italy}

% ...

%% file: acronyms.tex
\begin{acronym}
\acro{CUORE}{Cryogenic Underground Observatory for Rare Events}
\acro{NDBD}[0$\nu\beta\beta$ decay]{Neutrinoless Double Beta decay}
\acro{SM}{Standard Model}
\acro{NTD}{neutron transmutation doped thermistor}
\acro{FWHM}{full-width half-max}
\acro{MC}{Monte Carlo}
\acro{MCMC}{Markov-Chain Monte Carlo}
\acro{ROI}{region of interest}
\end{acronym}

\newcommand{\Q}{\ac{CUORE}\xspace}
\renewcommand{\NDBD}{\ac{NDBD}\xspace}
\newcommand{\SM}{\ac{SM}\xspace}
\newcommand{\NTD}{\ac{NTD}\xspace}
\newcommand{\FWHM}{\ac{FWHM}\xspace}
\newcommand{\MC}{\ac{MC}\xspace}
\newcommand{\MCMC}{\ac{MCMC}\xspace}
\newcommand{\ROI}{\ac{ROI}\xspace}

\newcommand{\detailtexcount}[1]{%
  \immediate\write18{texcount -merge -sum -q #1.tex output.bbl > #1.wcdetail }%
  \verbatiminput{#1.wcdetail}%
}

\newcommand{\quickwordcount}[1]{%
  \immediate\write18{texcount -1 -sum -merge -q #1.tex output.bbl > #1-words.sum }%
  \input{#1-words.sum} words%
}

\newcommand{\quickcharcount}[1]{%
  \immediate\write18{texcount -1 -sum -merge -char -q #1.tex output.bbl > #1-chars.sum }%
  \input{#1-chars.sum} characters (not including spaces)%
}

%% file: 01intro.tex
\section{Introduction}

Two-neutrino double beta (2\vbb) decay is a nuclear transition with the longest lifetime experimentally measured. This process occurs when two neutrons in a nucleus simultaneously decay emitting two anti-neutrinos and two electrons. This decay can act as background for a hypothetical process called neutrinoless double beta (0\vbb) decay ~\cite{Furry1939}, which may occur if the neutrino were a Majorana fermion~\cite{Vergados_2012}, and which violates lepton number conservation~\cite{FUKUGITA198645}. 0\vbb decay would be new physics and could explain the origin and nature of the neutrino mass states~\cite{RevModPhys.88.030501,RevModPhys.88.030502,Eguchi2003,PhysRevD.95.072006}.

Precision measurements of the 2\vbb decay half-life and studies of the 2\vbb decay spectral shape can provide important input for nuclear models \cite{Suhonen:2017krv,Simkovic:2018rdz,Moreno:2008dz,Domin:2004za}. Measurements are available in literature for the 2\vbb decay of various isotopes such as $^{116}$Cd (Aurora~\cite{Aurora.2nu.2018.PhysRevD}), $^{76}$Ge (GERDA~\cite{GERDA2015.2nu}), $^{100}$Mo (CUPID-Mo~\cite{CUPIDMo.2nu.precise}), $^{150}$Nd (NEMO-3~\cite{NEMO3.2nu.Nd.PhysRevD.94.072003}), $^{82}$Se (NEMO-3~\cite{NEMO-3:2019gwo}, CUPID-0~\cite{CUPID0.PhysRevLett.123.262501}), $^{136}$Xe (EXO-200~\cite{EXO200.2nu.Albert:2013gpz}, KamLAND-Zen~\cite{KamLANDZen2016, KamLANDZen2019.2nu}), and $^{96}$Zr (NEMO-3~\cite{NEMO3.2nu.Zr.NuclPhysA}). This paper will discuss the first measurement of the $^{130}$Te 2\vbb decay half-life performed with the unprecedented statistics of the \Q experiment.

The \Q experiment primarily searches for neutrinoless double beta decay (0\vbb) of \Tedecay\cite{CUOREPRL2017,CUOREPRL2020}, however other searches are possible~\cite{Q0.PhysRevC.97.055502, Q.IJMPA.33, CUORE0.excited.states.2019EPJC, CUORE.LowE.2017EPJC}.
This letter will outline first the \Q detector, the data collection, and the analysis of the 2\vbb decay of \isoTe. Thereafter, we provide a description of the technique used to fit the experimental data (comprised of events from both the 2\nbb decay and the background sources), and finally we present a discussion of the fit results.


\begin{figure*}[htbp]
  \centering
  \includegraphics[width=17.2cm]{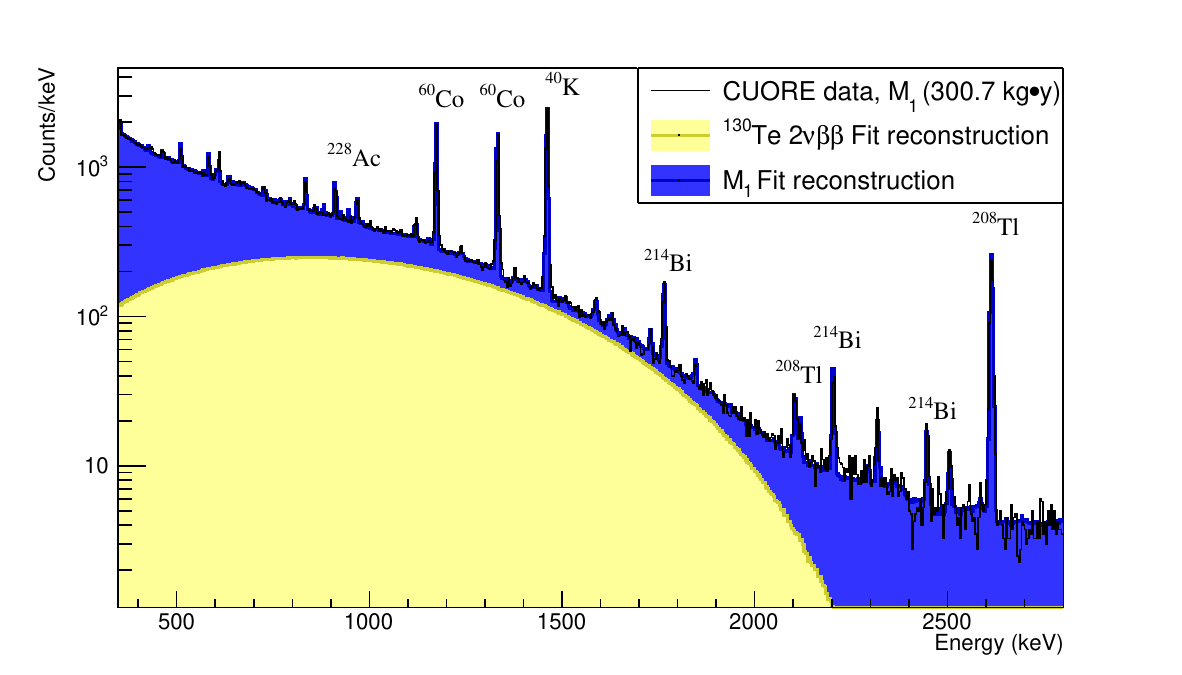}
  \caption{The observed \mone spectrum (\emph{black}) compared with its reconstruction as obtained by the background model (\emph{blue}). The reconstructed 2\nbb decay component is shown in \emph{yellow} for comparison. We observe that from 900 to 2000~keV more than 50\% of the \mone spectrum counts originate from the 2\nbb decay process. The experimental data and the spectra reconstructed by the fit have been converted back to 1~keV binning for illustrative purposes. Selected gamma lines from background contaminants are labeled.}
  \label{fig:Stacked}
\end{figure*}

%% file: 02detector.tex
\section{CUORE Detector}
CUORE is located underground at the Gran Sasso National Laboratory of INFN, Italy, with $\sim$3600~meter water equivalent overburden to shield from cosmogenic backgrounds~\cite{PhysRevD.58.092005,BELLINI2010169}. The \Q detector consists of 988 \TeO crystals \cite{Arnaboldi2010} arranged in 19 towers made of 13 floors of 4 crystals, read out as individual channels. Each crystal is 5$\times$5$\times$5 cm$^3$ in size and $\sim$750~g in mass, with a natural abundance of $\sim$34\% for \isoTe. The crystals are cooled to $\sim$10~mK at which point they have a low heat capacity and can be operated as cryogenic bolometers. The energy deposited in the crystal by particle interaction causes a temperature increase which is measured via a neutron transmutation doped (NTD) Ge thermistor~\cite{Haller1984}. The signal rise time is $\sim$100~ms, and the high energy resolution is second only to that of Ge detectors. Si heaters are used to inject regular reference pulses for a detector-based correction of thermal gain drifts from long term temperature variations.

The detector is housed in a large cryogen-free cryostat, cooled to $\sim$10~mK by a dilution refrigerator \cite{CUORE_cryostat}. The cryostat and the detector were constructed with strict radiopurity controls~\cite{Arnaboldi2010} in order to reduce the $\alpha$ and $\gamma$ backgrounds seen in \qino \cite{QINOPaper}, and further reduce the $\gamma$ background seen in \q \cite{Q0Detector}. The cryostat is equipped with two lead shields: a 6-cm thick shield of ancient Roman lead \cite{ALESSANDRELLO1998-Pb} at $\sim$4~K around and below the detector, and a 30-cm thick shield at $\sim$50~mK located above the detector. An additional external shield is comprised of a 25-cm thick layer of lead surrounded by a 20-cm thick layer of polyethylene. A layer (2~cm thick) of boric acid that absorbs thermalized neutrons is located between the two shields.

When 2\nbb decay occurs inside a bolometer the neutrinos escape without interacting, thus we detect the two electrons sum kinetic energy forming a continuous, $\beta$-like spectrum from 0~keV up to the Q-value of the decay (\Qbb = 2527~keV~\cite{Redshaw2009, 130Te.QValue.PhysRevC.80.025501, RAHAMAN2011412}). Background contributions to this spectrum originate from radioactivity in the detector and cryostat components. These backgrounds can be disentangled and quantified via careful analysis of the observed spectral shape and topological information in the segmented \Q detector in comparison to a detailed background model~\cite{Q0BackgroundRecon, CUPID0.BM}.

%% file: 03data.tex
\section{Data Collection}

\Q began taking data in early 2017. The data collected through mid 2019 are analyzed in this work and are grouped into \datasets datasets (physics data bounded by \maura{$^{232}$Th and $^{60}$Co} calibration data). The 2\vbb decay analysis requires high quality data over the whole energy range, specifically channels need to be both well-performing and well-calibrated. This led us to exclude 2 datasets from the analysis given that the large majority of channels did not satisfy these criteria. With this choice of dataset-channel we have \exposure\kgyr of \TeO exposure (\isoexposure\kgyr of \isoTe exposure).

The data itself are a collection of events, corresponding to a triggered waveform on a single bolometer. The modularity of the detector allows us to reconstruct the event topology via a time based coincidence analysis. Events are grouped into multiplets, $\mathcal{M}_{i}$ ($i$ = number of triggered bolometers) if they occur within a $\pm$\dtCoinc window on bolometers that are $\leq$ \drCoinc apart from each other, with a minimum energy of~ \dECoinc. Given the extremely low trigger rate of \Q bolometers (\updateValue{$\sim$1~mHz}) and the distance requirement, multiplets with $i>1$ contain practically no accidental coincidence events and are mainly induced by particles depositing energy in multiple crystals.

We split the data into three types of spectra: a multiplicity 1 (\mone) spectrum comprised of events where energy was deposited into a single bolometer, a multiplicity 2 (\mtwo) spectrum comprised of the single energies detected by each of the two bolometers simultaneously triggered, and a \msumtwo spectrum comprised of the sum energy of the \mtwo events. The energy of a 2\nbb decay event is deposited into a single bolometer with a probability obtained from Monte Carlo simulations of $\sim$ 90\%.~
The majority of backgrounds deposit energy across two or more bolometers (such as $\gamma$'s that scatter from one crystal into another or $\alpha$ decays that occur on a surface between two crystals), making the \mtwo and \msumtwo spectra useful for understanding backgrounds. Events with multiplicity higher than 2 are not considered in this analysis since they do not add new information. The specific steps of the data processing and selection criteria are found in \cite{CUOREPRL2020}. These steps include a dataset dependent evaluation of the energy calibration bias, and of the signal efficiency. The former, defined as the difference between the reconstructed peak position and its nominal value, is measured in calibration data for $\gamma$ lines that span from 511~\,keV to 2.6~\,MeV. The resulting bias is well below 0.5~\,keV for all the datasets. A similar result is obtained by fitting $\gamma$ peaks due to background sources. The signal efficiency, defined as the probability of a signal being triggered, assigned to a correct energy and multiplicity, and finally passing data selection cuts, has an energy dependent behavior and is asymptotic to $\sim$95\%.

%% file: 04specfit.tex
\section{Spectral Fit}
We analyze the events with energies from a threshold of \fitElo to \fitEhi, where the \mone spectrum is dominated by 2\nbb decay (between 900 to 2000~keV the contribution exceeds 50\% of total \mone events) along with $\gamma/\beta$ emissions from radioactive contaminants. To disentangle the 2\nbb decay signal we construct a background model (BM) that describes the data via a comprehensive list of possible sources. Guidelines for this work are taken from the \q BM~\cite{Q0BackgroundRecon} and the \Q background budget~\cite{CUOREBudget}. The background sources are radioactive contaminations located both in the bulk of the detector and cryostat components, on the surfaces of crystals, and materials with a line of sight to them. We also include cosmogenic muons. 

We developed a \GEANT4 \cite{Geant} \MC simulation \cite{CUOREBudget, Q0BackgroundRecon} which outputs the spectra produced by each source in the detector, reproducing all relevant features of the experimental data (e.g., multiplicity, time resolution, energy dependent trigger efficiencies, etc.). The elements of the \Q experiment, including the cryostat, are grouped into \numElements geometric entities used in the model as background source positions: the crystals, the copper structure holding the towers, the copper vessel enclosing the detector, the Roman lead, the internal and external shields made of modern lead, the cryostat thermal shields (grouped into two elements: the copper shields inside the Roman lead and those outside), and finally the internal lead suspension system. The BM uses \numSim simulated sources to fit the experimental data. One is the 2\nbb decay in the crystals, and 60 others refer to different contaminants in the \numElements elements listed above. These include bulk and surface $^{238}$U and $^{232}$Th contaminations (allowing for secular equilibrium breaks), bulk $^{60}$Co, $^{40}$K, and a few other long lived isotopes, as indicated in Fig.~\ref{fig:Stacked}. All these isotopes are identified from the presence of one or more characteristic $\gamma$ lines in the observed spectra. The only exception is $^{90}$Sr, a long-lived pure $\beta$ emitter, that could be present due to a hypothesized contamination by radioactive fallout. The remaining simulation (number 62) is the cosmogenic muon flux.
As described in~\cite{Q0BackgroundRecon} a variable binning is applied to all the spectra: the minimum bin size is \minBin, and bins with less than \minCounts counts are merged. All counts belonging to a single $\gamma$ line are combined into a single bin to avoid systematics from the modeling of the $\gamma$ peak shapes in MC simulations.
Finally, the trigger efficiency vs. energy and the efficiency of quality cuts are included in the analysis as global parameters.

The observed spectrum is reconstructed by simultaneously fitting a linear combination of the \numSim \MC simulated spectra to the \mone, \mtwo, and \msumtwo data. The fit is done with a Bayesian approach using a \MCMC, implemented in the JAGS software package, to sample the joint posterior probability density function (PDF) of the fit parameters~\cite{JAGS1, JAGS2, JAGS3, JAGS_Proceeding}. The likelihood is a product, over bins and spectra, of Poisson distributions that give the probability of drawing the experimental counts as a function of the MC spectra normalizations.
To prevent bias while tuning data quality cuts and setting the fitting procedure, the \MC normalization coefficient was blinded to keep the extracted 2\vbb decay half-life in terms of a nonphysical ratio that could not be compared to previous results.

For each source, except cosmogenic muons, a uniform prior is used (the activity can span from zero to 10 times the maximum activity compatible with the measured spectrum). For muons, additional information is gained from the high multiplicity spectra (\multi $>$ \muonMulti) where muons become dominant, and is used to extract a Gaussian prior for the BM fit. The fit result is a joint posterior PDF for the \numSim parameters from which we extract the marginalized posterior PDF for the 2\vbb decay rate. The fitting procedure closely follows the description in~\cite{Q0BackgroundRecon} and a paper detailing the \Q BM is in preparation.

%% file: 05bmsystematics.tex
\section{Model Systematics}

The background model is able to reproduce the major features of the observed spectra (see Fig.~\ref{fig:Stacked}) with a global \chisq/d.o.f of \chiglobal/\dofglobal ($\chi^{2}_{\rm red}$ = \redchiglobal). The sub-optimal agreement between the data and the \MC likely arises from an imperfect modelling of source position and distribution. Increased statistics from more data will allow for refinement of the background model by better identifying source locations or additional sub-dominant contaminants. This fit makes very limited use of the data and it is based on a simplified description of sources, therefore the result is not particularly informative on the specific position and intensity of a source. Overall the background composition matches very well the expectation discussed in~\cite{CUOREBudget} with a few exceptions. We see an excess of $^{238}$U in the cryostat elements. The localization is not clear, but the 2\nbb decay result is insensitive to the source position. We also have a quite evident $^{210}$Pb surface contamination of the copper of the tower holding structure that is $\sim$100 times higher than in CUORE-0. Its major contribution to the measured spectra is the $\alpha$ peak at 5.3 MeV due to $^{210}$Po. The shape of the peak proves that the contamination is right at the surface of the copper, likely due to $^{222}$Rn exposure. We observe an excess in the $^{60}$Co crystal contamination compared with the expected 1 nBq/kg \cite{CUOREBudget}. However, this is anti-correlated with $^{60}Co$ in the copper of the tower holding structure. Increased statistics and a more extensive use of coincidences will allow to clarify this point.



\begin{figure}[htbp]
  \centering
  \includegraphics[width=8.6cm]{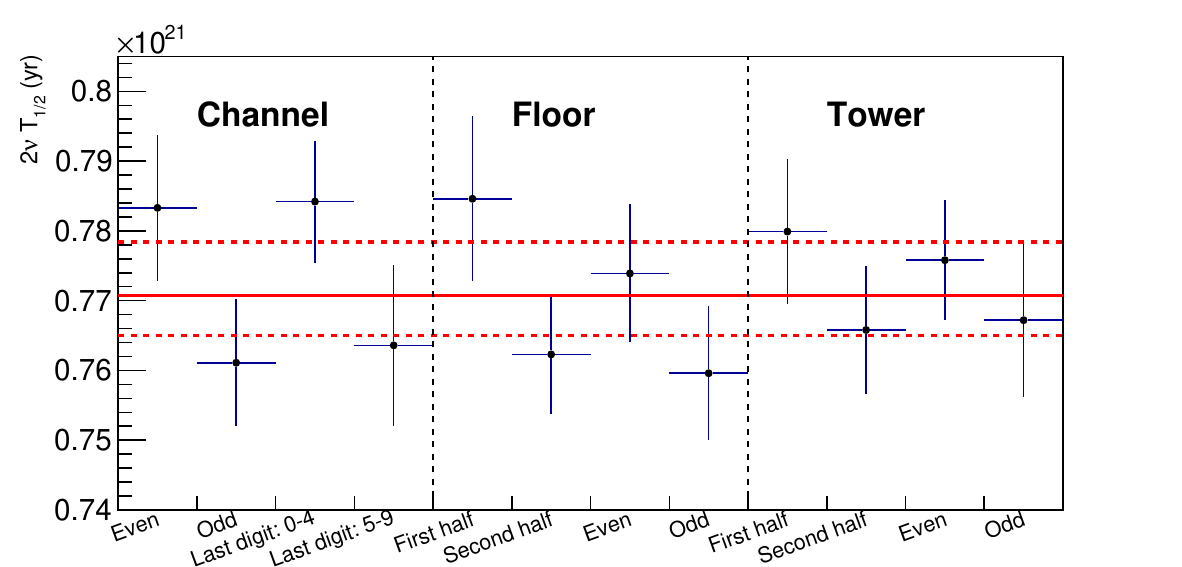}
  \caption{Results of the 2\nbb decay half-life from the tests in which data are fit by splitting the detector in half in different ways \emph{Channel:} Detector split based on channel numbers assigned to the individual crystals (even, odd, last digits: 0-4, 5-9). \emph{Floor:} Detector split based on tower floor number (first and second halves, even, odd). \emph{Tower:} Detector split based on tower number (first and second halves, even, odd). The results are compared to the one obtained in the full statistics fit (red solid line) and its statistical uncertainty range (red dashed lines)}
  \label{fig:Systematics}
\end{figure}

In order to check the stability of the 2\nbb decay half-life result we run multiple fits over the whole dataset varying aspects of the background model. In particular we test two different models for the 2\nbb decay spectral shape, we alter the list of background sources used and we remove the $^{90}$Sr source. As additional probes of our sensitivity to various aspects of the BM sources we fit subsets of data in which we split the detector in half in different ways (see Fig.~\ref{fig:Systematics}) and perform the fit on single datasets. 

\begin{itemize}
    \item[--]{\textbf{2\nbb Decay Model}}
    There are two competing models for 2\nbb decay, yielding slightly different spectral shapes~\cite{Moreno:2008dz,Domin:2004za,Barea:2013bz}. The Single State Dominance (SSD) is the default used in this analysis. It results in a better fit quality, and might be the first experimental hint for SSD dominance in $^{130}$Te 2\nbb decay. The alternate mechanism, Higher State Dominance (HSD), yields a slightly worse fit and a few percent increase of the 2\nbb half-life. As this is the only shape test we perform on the 2\nbb decay spectrum, we conservatively assume this systematic to be double-sided and estimate the uncertainty as 68\% of the difference between the fits with HSD and SSD: $\pm$1.3\%.

    \item[--]{\textbf{Energy Threshold}}
    The energy threshold used in this analysis is \fitElo. If we vary this threshold in the range of 300--800~keV we observe an increase of the 2\nbb decay rate. We assume this systematic to be uniform between the best fit and the value that deviates the most. We symmetrize this uncertainty around the best fit to account for possible deviations given by untested threshold values, with a result of $\pm$0.4\%.

    \item[--]{\textbf{Geometrical effect}}
    All contaminants in the model are uniformly distributed in the 9 simulated elements. To investigate possible biases we compare fits done by splitting the detector according to crystal, floor, or tower number (see Fig.~\ref{fig:Systematics}). Each pair of results are statistically compatible with each other to within two sigma. We take the pair with the largest splitting, subtract the statistical error and interpret the result as the 1$\sigma$ uncertainty of a flat distribution: $\pm$0.8\%
    
    \item[--]{\textbf{$^{90}$Sr}}
    As mentioned, a background source due to possible crystal contamination with $^{90}$Sr was introduced. This is the only long-lived pure $\beta$ emitter produced by fission that produces a background (via its daughter $^{90}$Y) extending up to 2.2~MeV, without any associated gamma emission that would allow us to constrain its activity~\cite{CUPID0.PhysRevLett.123.262501}. The 2\vbb decay result is weakly sensitive to this contaminant, as upon its removal the counts ascribed to 2\vbb decay increase resulting in a slightly shorter half-life. Since $^{90}$Sr has no clear signature we use it as a proxy for the removal or addition of components to the BM. We take the systematic to be symmetric and 68\% of the difference between the best fit and the fit without, giving $\pm$0.3\%.
    
    \item[--]{\textbf{Datasets}}
    We investigated the 2\vbb decay result stability in time by fitting separately each of the 5 datasets used in this analysis and observed only statistical variations in the fit result. We also fit the two excluded datasets and use the result to quantify the bias introduced by their removal. This yields an asymmetric uncertainty of +0.3\% and -1.1\%.
    
    \item[--]{\textbf{List of background sources}}
    In the reference fit we use \numSim sources, selected to be comprehensive, as extracted from the CUORE-0 background model~\cite{Q0BackgroundRecon}. Given the limited statistics used in this work and the choice of fitting only the $\gamma$ region some of the sources could be degenerate with each other while others cannot be identified easily.
    To check how the fit performs with a different background source list, all components with contributions compatible with 0, including $^{90}$Sr, are removed. This reduces the number of distinct background source components down to \numBgListSyst. This has an impact on the 2\nbb decay fit result compatible to that resulting from the removal of the $^{90}$Sr alone.
    
    
\end{itemize}

As a result of this study we can conclude that all the systematics we explored are at most in the range of $1\%$. The dominant contribution comes from the uncertainty in the decay model (SSD vs HSD) which may be improved with increased statistics or theoretical input. Finally, other sources of uncertainties such as the efficiency of our coincidence selection, the chance of mixing up \mone with \mtwo events, or the efficiency of pulse shape cuts, have an overall impact on the final error that is lower than 0.1\%. The Monte Carlo statistics, though optimized to yield negligible error, is properly accounted for in the fitting procedure.

%% file: 06results.tex
\section{\texorpdfstring{2\nbb}{2vbb} Decay Results and Discussion}

To extract a robust estimate of the 2\nbb decay half-life, we combine our systematics in quadrature. Through the unblinding of the correct normalization coefficient for the MC spectrum, we obtain the measurement of the 2\nbb decay half-life of \isoTe. Though the posterior for $^{90}$Sr is compatible with null activity, the insertion of $^{90}$Sr in the BM does weakly distort the 2\nbb decay posterior. Removal of the $^{90}$Sr source results in a symmetric posterior, however we choose to include this source due to the high anti-correlation with 2\nbb decay.

We use an isotopic abundance of \isoTe of (34.167 $\pm$ 0.002)\%~\cite{FEHR200483}. From this, and the systematic uncertainties described previously, we obtain a half-life of \hlifeResult, a value consistent with previous measurements (see Table~\ref{tab:halflives}). This result is the most precise measurement of the 2\nbb decay half-life of \isoTe to date and one of the most precise measurements of a 2\nbb decay half-life. It represents a substantial improvement over previous measurements from NEMO-3 \cite{NEMO3Te2nu} and \q \cite{Q0BackgroundRecon} owing to the \Q strict radiopurity controls, the improved signal-to-noise ratio, the increased statistics, and the robust background model.



\begin{table}[htbp]
  \centering
  \caption{Chronology of $T_{1/2}^{2\nu}$ measurement in \isoTe. The relative uncertainty refers to statistical and systematic errors summed in quadrature.}
  \label{tab:halflives}
  \begin{tabular}{lccc}
  \hline
  \hline
  & $T_{1/2}^{2\nu}$ ($10^{20}$~yr) & Relative Uncert. & Ref. \\
  \hline
  MiBeta & $6.1\pm1.4 ~{}^{+2.9}_{-3.5}$& 57\% & 2003 \cite{MiDBD2nu} \\
  NEMO-3 & $7.0 \pm 0.9 \pm 1.1$ & 20\% & 2011 \cite{NEMO3Te2nu} \\
  CUORE-0 & $8.2\pm0.2\pm0.6$ & 7.7\% & 2016 \cite{Q0BackgroundRecon}\\
  CUORE  & $\hlife^{\hlifestatUp\hlifesystUp}_{\hlifestatLow\hlifesystLow}$ & \hlifefracuncert\% & (this result) \\
  \hline
  \hline
  \end{tabular}  
\end{table}


%% file: 07conclusion.tex
\section{Conclusion}
In this paper we described the analysis of the \isoTe 2\nbb decay measured with CUORE. We exploit the geometry of the \Q detector to tag single scatter and multiple scatter events to obtain separate spectra dominated by 2\nbb decay and background events, respectively. The \isoTe 2\nbb decay half-life is measured to be \hlifeResult. Compared to previous results (Table~\ref{tab:halflives}) this is the most precise determination of the 2\nbb decay half-life in \isoTe. The present result is dominated by a $\sim$2\% systematic uncertainty. Further improvement will require a better understanding of the background sources localization, as indicated by the observed BM variations with different geometrical detector splittings. This refinement, as well as an improved study of the SSD vs HSD models, is feasible in the near future given the increased statistics being collected by CUORE.


%% file: 08acknowledgments.tex

\section{Acknowledgments}

The CUORE Collaboration thanks the directors and staff of the Laboratori Nazionali del Gran Sasso and the technical staff of our laboratories. 
This work was supported by the Istituto Nazionale di Fisica Nucleare (INFN); the National Science Foundation under Grant Nos. NSF-PHY-0605119, NSF-PHY-0500337, NSF-PHY-0855314, NSF-PHY-0902171, NSF-PHY-0969852, NSF-PHY-1307204, NSF-PHY-1314881, NSF-PHY-1401832, and NSF-PHY-1913374; and Yale University. 
This material is also based upon work supported by the US Department of Energy (DOE) Office of Science under Contract Nos. DE-AC02-05CH11231 and DE-AC52-07NA27344; by the DOE Office of Science, Office of Nuclear Physics under Contract Nos. DE-FG02-08ER41551, DE-FG03-00ER41138, DE-SC0012654, DE-SC0020423, DE-SC0019316; and by the EU Horizon2020 research and innovation program under the Marie Sklodowska-Curie Grant Agreement No. 754496.  
This research used resources of the National Energy Research Scientific Computing Center (NERSC).
This work makes use of both the DIANA data analysis and APOLLO data acquisition software packages, which were developed by the CUORICINO, CUORE, LUCIFER and CUPID-0 Collaborations.

%% file: 10suppl.tex
\section{Supplemental Materials}


Here we present additional plots to illustrate key results from the \Q background model fits with respect to the 2\nbb decay half-life result. In particular we show the fit results for the 3 spectra used in the fits (\mone, \mtwo, \msumtwo) which show good agreement between data and model. We also include a comparison of the 2\nbb decay posterior with and without the $^{90}$Sr source to illustrate that, while there is some slight distortion, the overall impact is quite negligible. This is further supported by the posterior of the $^{90}$Sr contribution.

\begin{figure*}[htbp]
  \centering
  \includegraphics[width=17.2cm]{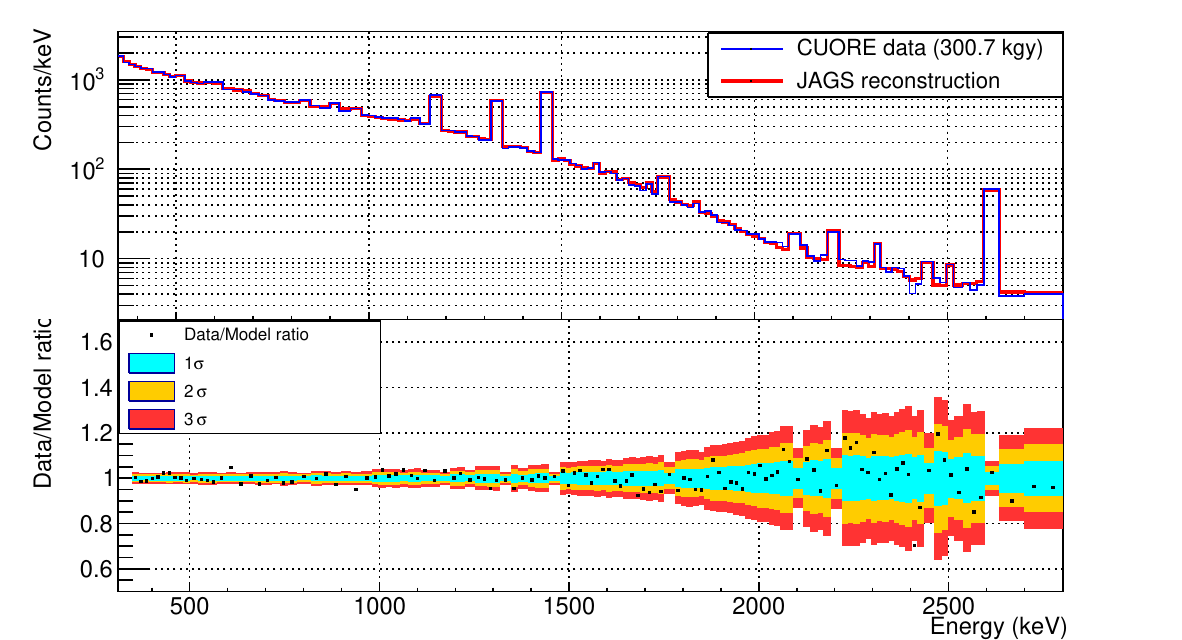}
  \caption{Top: The measured \mone spectrum (\emph{blue}) and its
    reconstruction (\emph{red}). The spectra are binned with an
    adaptive binning to contain peaks into a single bin (to avoid
    dependence on the peak shape), while also achieving good
    resolution of the continuum shape. Bottom: The ratio of the data to the reconstructed model with 1$\sigma$, 2$\sigma$ and 3$\sigma$ error bars. It is clear from the data that we are able to faithfully reconstruct the continuum and peaks from sources.}
  \label{fig:M1L0}
\end{figure*}

The \mone spectrum (Fig.~\ref{fig:M1L0}) is comprised of single-crystal events which contain a significant contribution from 2\nbb decay. The fit residuals show that in the region of 1-2~MeV the reconstructed spectrum matches the observed data quite well.

\begin{figure*}[htbp]
  \centering
  \includegraphics[width=17.2cm]{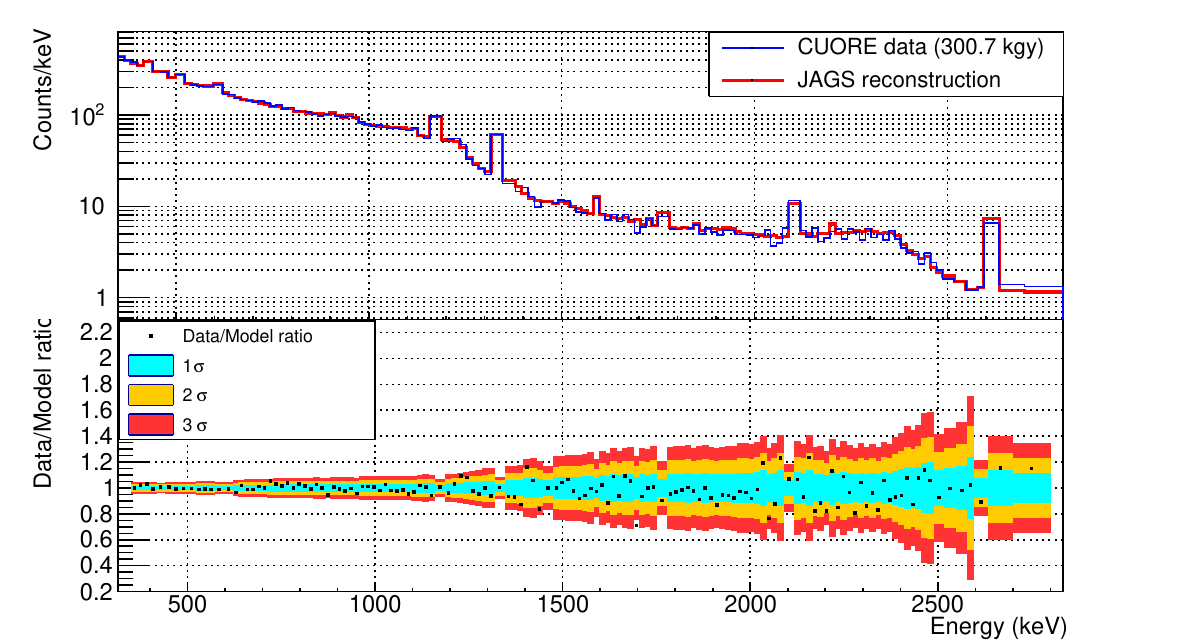}
  \caption{The \mtwo spectrum provides an indication of the localization of sources as they are populated by coincident events from neighboring crystals. Top: The measured \mtwo spectrum (\emph{blue}) and its
    reconstruction (\emph{red}). The spectra are binned with an adaptive binning to contain peaks into a single bin (to avoid dependence on the peak shape), while also achieving good resolution of the continuum shape. Bottom: The ratio of the data to the reconstructed model with 1$\sigma$, 2$\sigma$ and 3$\sigma$ error bars.}
  \label{fig:M2}
\end{figure*}

\begin{figure*}[htbp]
  \centering
  \includegraphics[width=17.2cm]{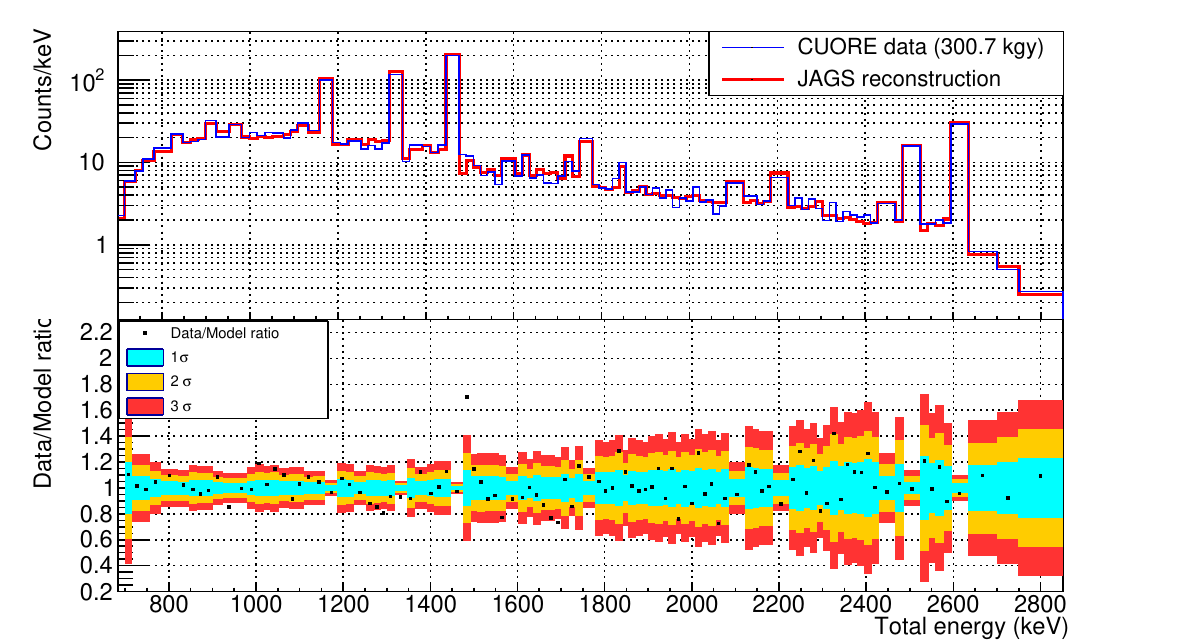}
  \caption{The \msumtwo spectrum better indicates the contributions to the background that originate from various $\gamma$ lines as a result of interactions with two crystals. Top: The measured \msumtwo spectrum (\emph{blue}) and its
    reconstruction (\emph{red}). The spectra are binned with an adaptive binning to contain peaks into a single bin (to avoid dependence on the peak shape), while also achieving good resolution of the continuum shape. Bottom: The ratio of the data to the reconstructed model with 1$\sigma$, 2$\sigma$ and 3$\sigma$ error bars.}
  \label{fig:M2sum}
\end{figure*}

In Figs.~\ref{fig:M2} and \ref{fig:M2sum} we show two views of the \mtwo data: a spectrum from the individual components of the \mtwo multiplets (i.e. the \mtwo spectrum), and a spectrum from the sum of the two components (i.e., the \msumtwo spectrum).
The \mtwo spectrum shown in Fig.~\ref{fig:M2} displays the energy spectra from individual components of \mtwo multiplets. Events in the \mtwo spectrum provide useful information on the localization of contaminations,  as the strict coincidence criterion described earlier limits these to the nearest neighbor crystals.
The \msumtwo spectrum in Fig.~\ref{fig:M2sum} shows the summed energy of each member of an \mtwo multiplet. The \msumtwo events show more clearly the $\gamma$ lines that produce the physical interactions in two crystals. We see here, that in the $\gamma$ region the background model accurately reconstructs the observed spectrum. Minor disagreements in some of the peaks is attributed to a potential for further improvement in source localization throughout the detector.

\begin{figure*}[htbp]
  \centering
  \includegraphics[width=17.2cm]{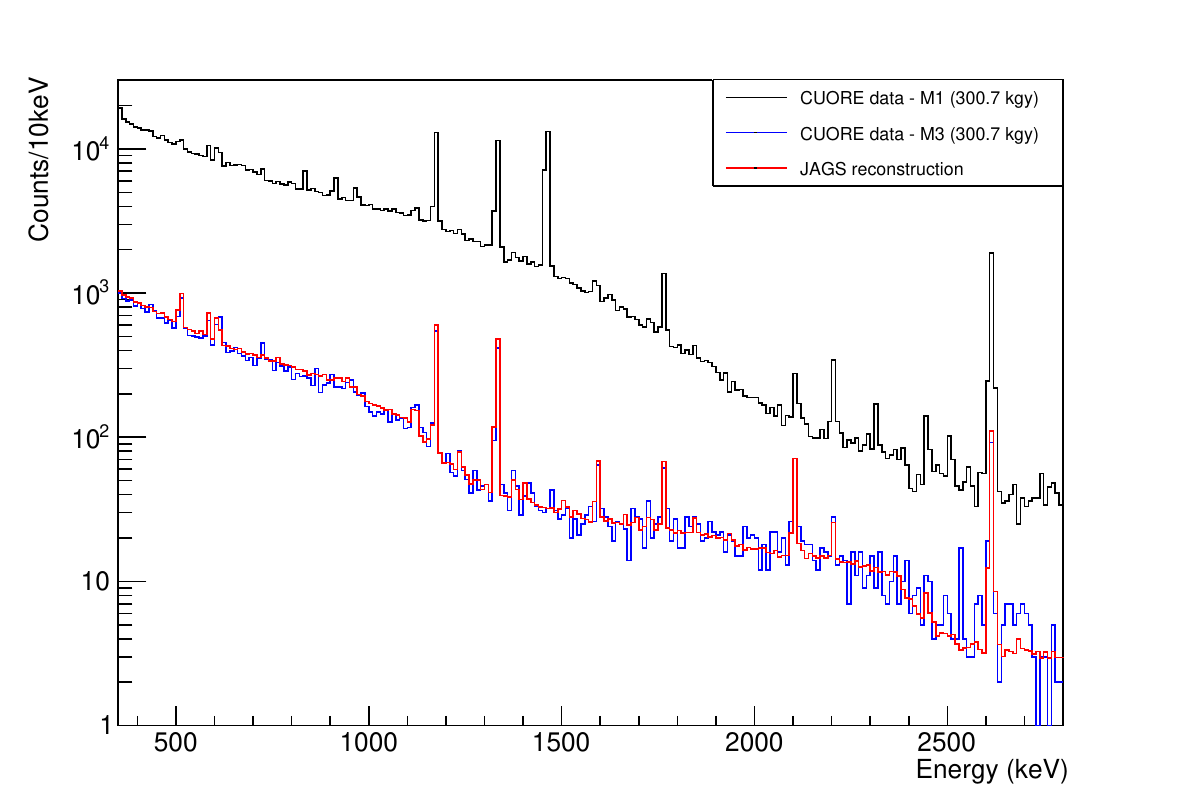}
  \caption{Spectrum of events with multiplicity three shown within the fit range. Though these data are not used in the fit, they are well reconstructed.}
  \label{fig:M3}
\end{figure*}

The reconstruction of the spectrum of events with multiplicity 3 (not used in the fit) is shown in Fig.\ref{fig:M3}. The blue histogram is the spectrum of experimental data, the red histogram its reconstruction according to the background model. The black histogram is the \mone spectrum, which is used as reference.

\begin{figure*}
  \centering                                   \hfill
  \begin{minipage}{8.6cm}
    \subfigure[]{\label{fig:posterior2nu}\includegraphics[width=8.6cm]{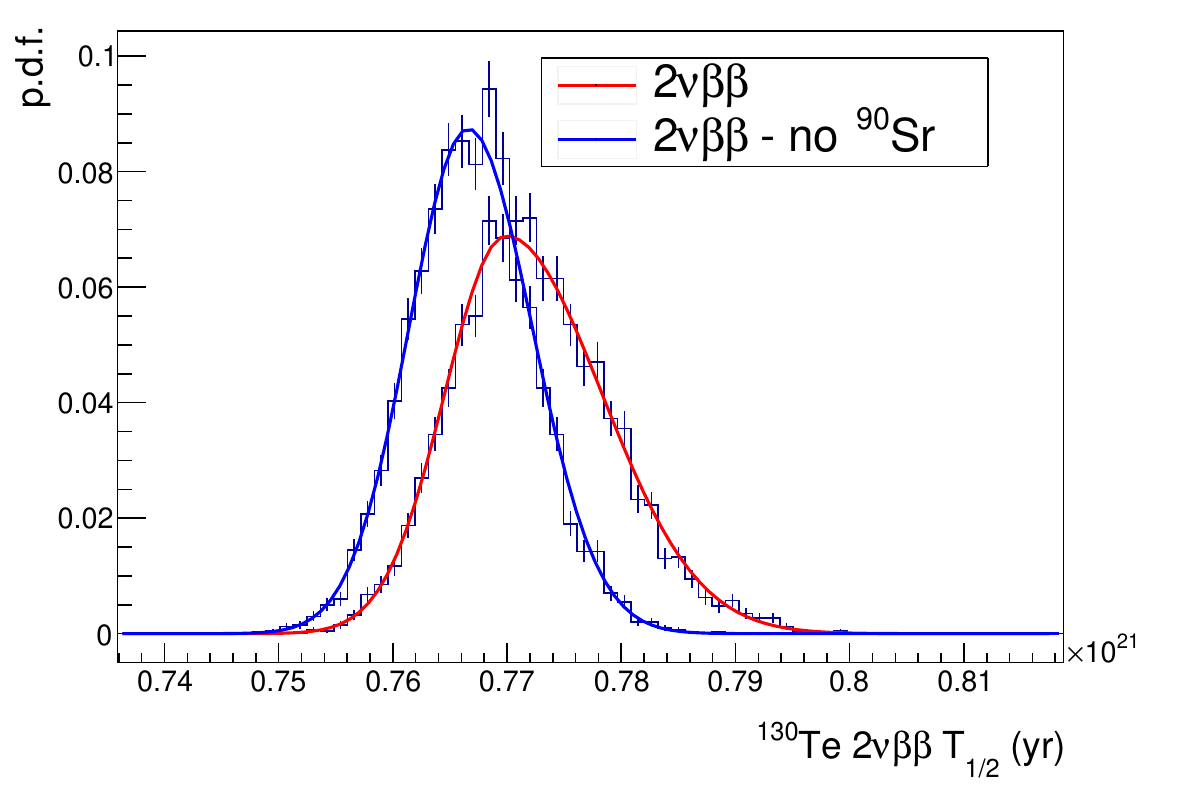}}
  \end{minipage}
  \begin{minipage}{8.6cm}
    \subfigure[]{\label{fig:90Sr}
      \includegraphics[width=8.6cm]{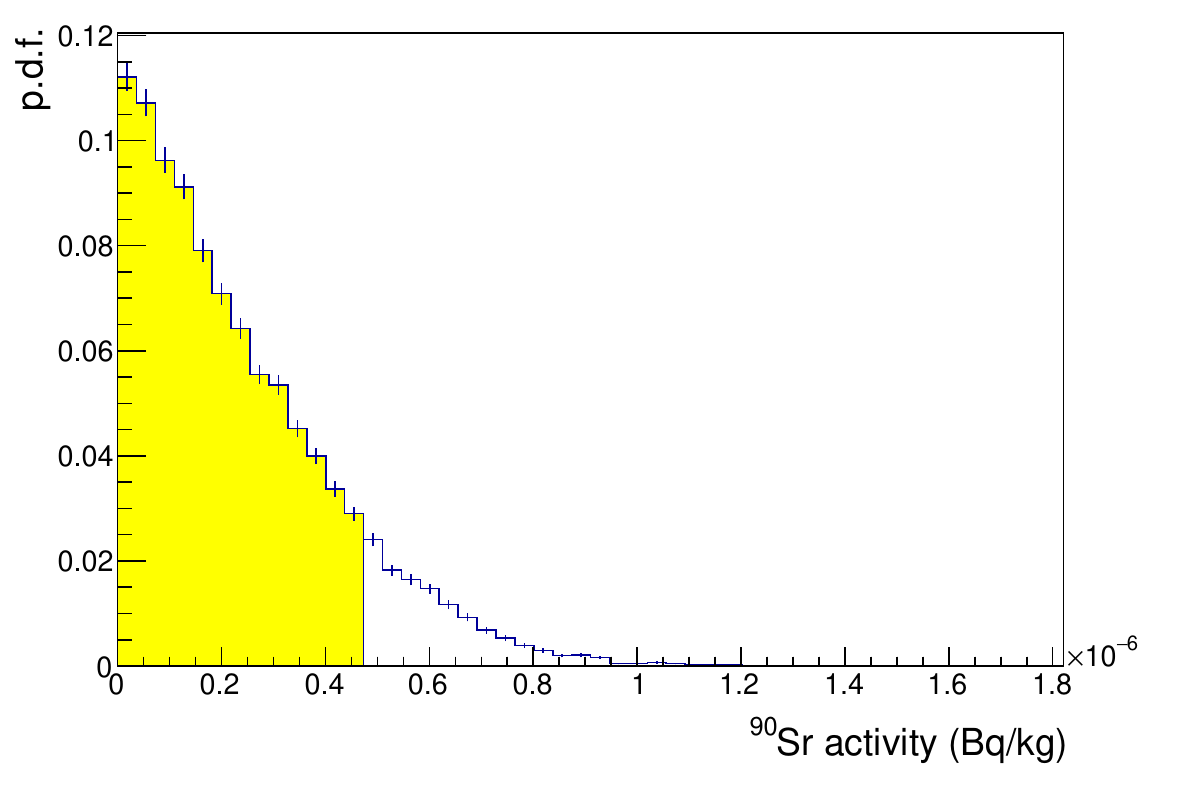}%
    }
    \hspace{1cm}
  \end{minipage}
  \hfill
  \caption{(a) Posterior of the 2\nbb decay reference fit (red line)  compared with the posterior of the 2\nbb decay fit with the $^{90}$Sr source turned off (blue line). The latter posterior is far more symmetric due to the anti-correlation between the two sources causing a distortion in the reference fit. (b) Posterior for the contribution from $^{90}$Sr in the reference fit, with the 90\% C.I. in yellow. The posterior peaks at a value consistent with 0 activity, indicating that the contribution from this source on the 2\nbb decay half-life measurement is negligible. Since there is a slight anti-correlation and distortion of the 2\nbb posterior we make a conservative choice to include it in the model for the 2\nbb decay fit result.}
  \label{fig:posteriors}
\end{figure*}

There is a possible contribution to the background from a fission product, $^{90}$Sr. This is a pure $\beta$ emitter with a decay energy of 0.564~MeV to $^{90}Y$, which is another nearly pure $\beta$ emitter with a 2.28~MeV endpoint allowing the decay chain to contribute to the background up through this energy. The net effect is an anti-correlation in the fit between the $^{90}$Sr rate and the 2\nbb decay rate. Systematics checks (described earlier) show that toggling this potential source off alters the 2\nbb decay rate by $\sim$0.6\%. The inclusion of the $^{90}$Sr source causes a slight asymmetry in the 2\nbb decay posterior (Fig.~\ref{fig:posterior2nu}). By fitting this distribution with a 2-sided Gaussian we get the left and right uncertainty range. Without $^{90}$Sr the posterior is far more symmetric requiring only a single Gaussian to fit. The half-life from this analysis with the $^{90}$Sr source is $\hlife^{\hlifestatUp}_{\hlifestatLow}$ $\times10^{20}$~yr, compared to the result without the source: \hlifeNoSr $\pm$\hlifestatNoSr $\times10^{20}$~yr. As there is an anti-correlation between the 2\nbb and $^{90}$Sr decay rates we keep this source in the final fit to get a conservative result. An examination of the $^{90}$Sr posterior itself (Fig.~\ref{fig:90Sr}) shows that this is a conservative approach as the posterior indicates a most probable value of zero.

\begin{figure*}
  \centering                                   \hfill
  \begin{minipage}{8.6cm}
    \subfigure[]{\includegraphics[width=8.6cm]{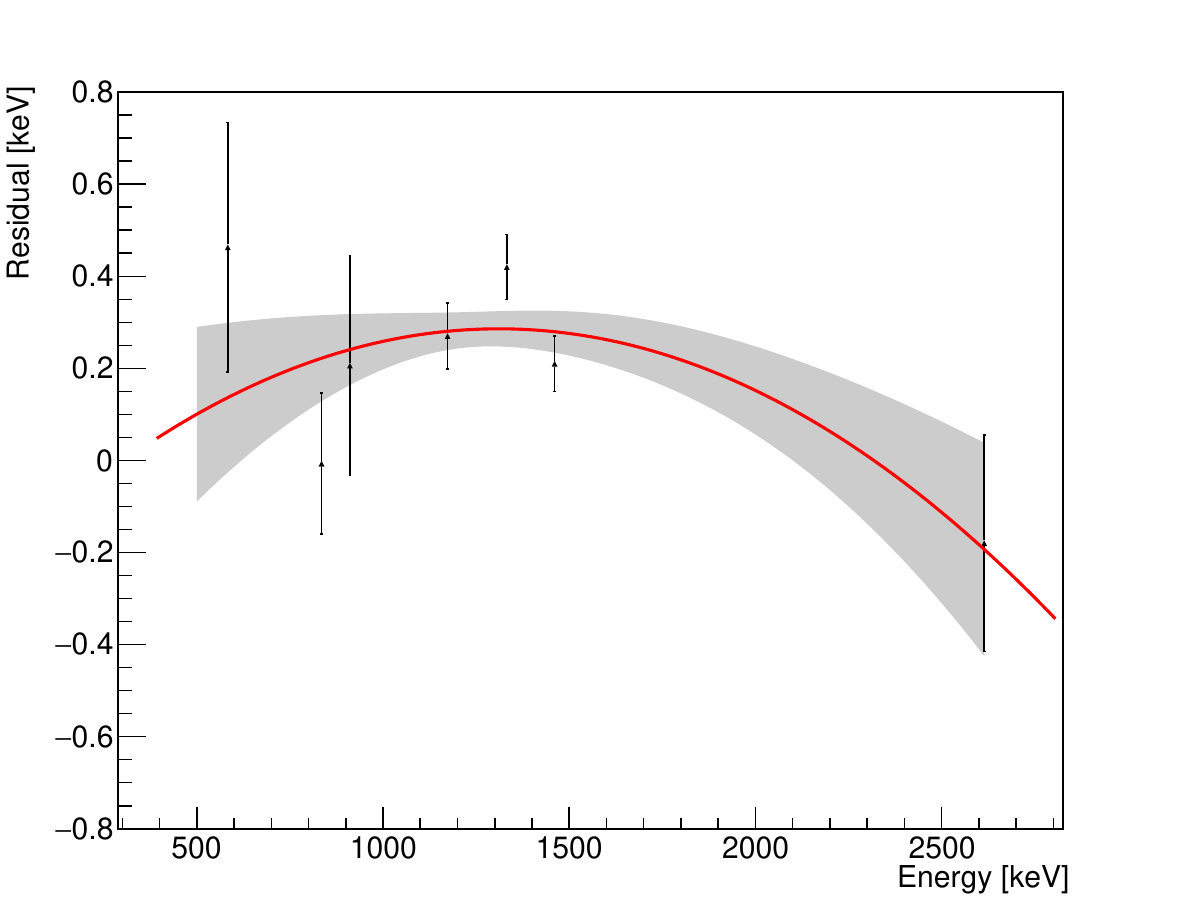}}
  \end{minipage}
  \begin{minipage}{8.6cm}
    \subfigure[]{
      \includegraphics[width=8.6cm]{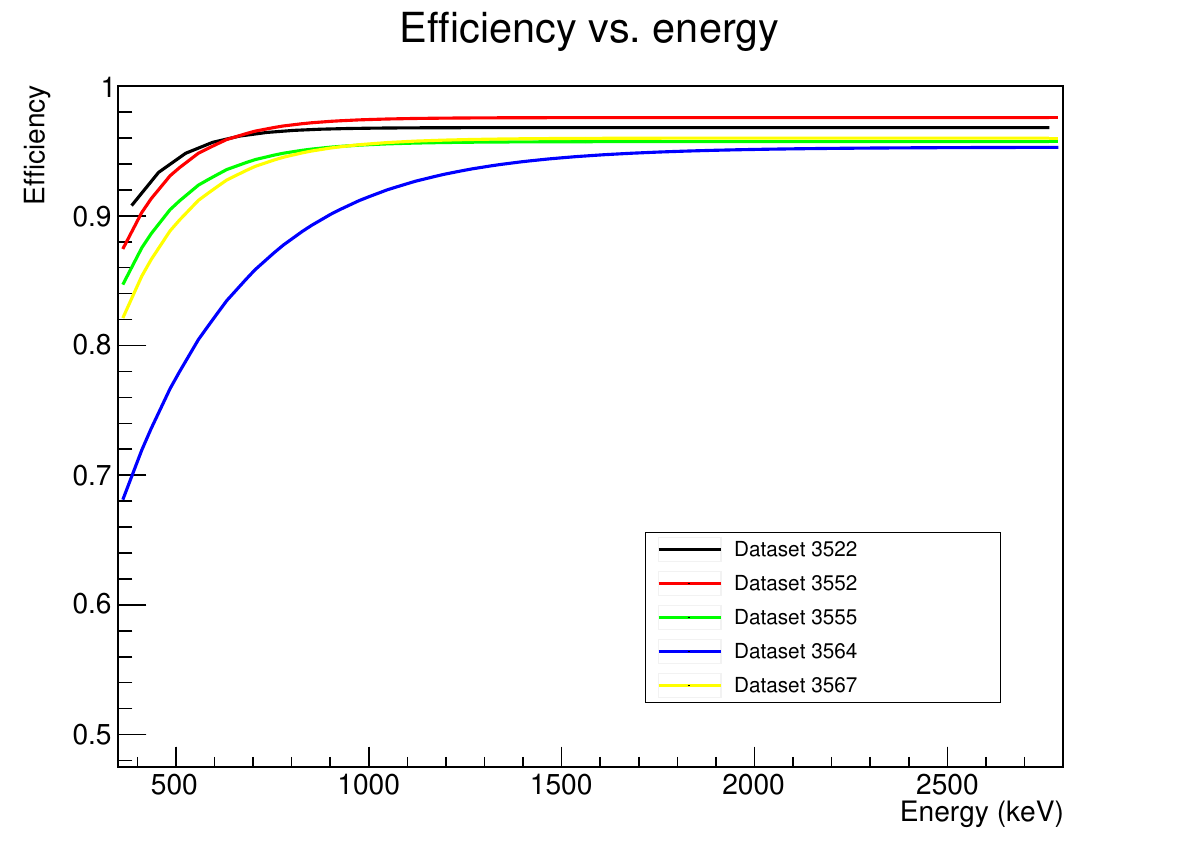}%
    }
    \hspace{1cm}
  \end{minipage}
  \hfill
  \caption{(a) Example of the energy bias as measured in a dataset: the plot shows the difference between a reconstructed peak position and its nominal value for some of the more intense $\gamma$ lines observed in a dataset. (b) Signal efficiency, defined as the probability of a signal being triggered, assigned to a correct energy and multiplicity, and finally passing data selection cuts, as obtained fitting experimental data. The efficiency approaches a constant value as energy increases but in some datasets the convergence to the asymptotic value is faster than in others.}
  \label{fig:technical}
\end{figure*}

We close this section with two more technical plots that show the energy calibration bias (Fig.\ref{fig:technical}(a)) of a typical dataset and the signal efficiency of our data selection criteria vs. energy. (Fig.\ref{fig:technical}(b))

